\documentclass[11pt,letterpaper]{article}

\usepackage[letterpaper,margin=1in]{geometry}
\usepackage{newtxtext,newtxmath} 
\usepackage{amsmath}
\usepackage{graphicx}
\usepackage{array}
\usepackage{booktabs}
\usepackage{caption}
\usepackage{cite}
\usepackage{enumitem}
\usepackage{titlesec}
\usepackage{changepage}
\usepackage{url}
\usepackage{tabularx}
\usepackage{graphicx}
\usepackage{subcaption}

\titleformat{\section}
  {\centering\bfseries\fontsize{12}{14}\selectfont}
  {}{0pt}{}
\titlespacing*{\section}{0pt}{12pt}{12pt}

\titleformat{\subsection}
  {\bfseries\fontsize{12}{14}\selectfont}
  {}{0pt}{}
\titlespacing*{\subsection}{0pt}{12pt}{6pt}

\titleformat{\subsubsection}
  {\bfseries\fontsize{12}{14}\selectfont}
  {}{0pt}{}
\titlespacing*{\subsubsection}{0pt}{12pt}{6pt}

\newenvironment{ansabstract}
  {\begin{adjustwidth}{0.5in}{0.5in}\fontsize{10}{12}\selectfont}
  {\end{adjustwidth}}

\newcommand{\corrauthornote}{Email: jiezhang@utdallas.edu}

\begin{document}

\begin{center}
{\bfseries\fontsize{14}{16}\selectfont
Load-Following Dynamics of an SMR-Powered All-Electric Ship with Battery Energy Storage
\par}

\vspace{6pt}

{\bfseries
Ali Mahboub Rad and Jie Zhang\footnotemark[1]\par}
\vspace{6pt}

The University of Texas at Dallas, Richardson, TX 75080, USA 
\end{center}
\footnotetext[1]{\corrauthornote}

\vspace{12pt}

\begin{center}
\end{center}

\section*{ABSTRACT}
\begin{ansabstract}
Small modular reactors (SMRs) are being considered as compact, high-energy-density power sources for future all-electric ships. In such systems, propulsion-load variations can propagate through the electrical network, turbine-generator set, steam generator, and reactor, producing coupled nuclear, thermal, mechanical, and electrical transients. This paper presents an integrated dynamic load-following analysis of an SMR-powered all-electric ship with and without battery energy storage system (BESS) support. The model couples an integral pressurized water reactor, a once-through moving-boundary steam generator, a simplified turbine model, a synchronous generator, an AC shipboard electrical network, an asynchronous propulsion motor, and a grid-connected BESS. A propulsion-load reduction is applied through the motor torque command, and the responses of the SMR-alone and SMR+BESS configurations are compared. The results show that, without BESS support, the sudden decrease in propulsion motor power creates a temporary surplus between turbine mechanical power and electrical demand, causing a large over-frequency transient and rapid steam-cycle adjustment. With BESS support, the battery absorbs excess electrical power during the transient, reducing the frequency peak from approximately 61.1 Hz to about 60.06 Hz. The BESS also indirectly smooths the upstream steam-cycle and reactor responses, including steam flow, steam-generator pressure, feedwater flow, coolant temperature, reactivity, reactor thermal power, and steam-generator region lengths. These results demonstrate the importance of modeling nuclear-powered electric ships as coupled multi-domain systems and show that BESS integration can improve short-term load-following performance.

\textbf{Keywords:} small modular reactor; all-electric ship; load following; battery; integrated energy system

\end{ansabstract}


\section*{INTRODUCTION}

The maritime sector is increasingly seeking high-power, reliable, and low-carbon energy sources for future ship propulsion and onboard electric loads. All-electric ship architectures are attractive because propulsion motors, ship-service loads, auxiliary systems, and future mission loads can be supplied through a common electric power system. Small modular reactors (SMRs) are a potential prime mover for such applications because of their high energy density, compact configuration, long refueling interval, and ability to provide continuous power for long-duration operation \cite{MaritimeSMRReview,ABSINLNRICReport,Holbert2025MaritimeReactors}.

Load-following capability is especially important for nuclear-powered electric propulsion. In all-electric ships, propulsion motor demand may change rapidly during acceleration, deceleration, maneuvering, direction changes, or operation under varying sea and weather conditions \cite{Badakhshan2023SMRBESSAES,Badakhshan2025SMRBESSSizing}. In an SMR-powered all-electric ship, these propulsion changes first appear as electrical load variations at the generator terminals. The resulting power imbalance can affect generator electrical output and system frequency, and the disturbance can then propagate upstream through turbine mechanical power, steam flow, steam-generator heat removal, primary-loop temperature, reactivity feedback, and reactor thermal power. Therefore, propulsion-load changes should be evaluated as coupled nuclear, thermal, mechanical, and electrical transients rather than as isolated motor-load events.

Previous studies have investigated SMR-based all-electric ships from several perspectives, including battery energy storage coupling, frequency support, reliable shipboard power system operation, medium-voltage DC networks, thermal-electrical coupling, energy dispatch, and SMR/battery sizing \cite{Badakhshan2023SMRBESSAES,Badakhshan2025SMRBESSSizing,MVDCThermalElectricShip}. These studies provide important insight into shipboard electrical stability and energy management strategies. However, many of them primarily emphasize electrical variables, storage response, voltage/frequency regulation, or planning-level operation, while the internal reactor and steam-cycle responses during propulsion-load-following events are not the main focus.

Integrated nuclear-electric propulsion studies have shown the importance of representing the reactor, turbine-generator, electrical distribution system, propulsion motor, and energy storage as coupled dynamic subsystems \cite{NuclearSubmarineHIL,SMRSubmarineDigitalTwin}. In parallel, studies of helical-coil once-through steam generators have shown that steam-generator dynamics can be important during changes in flow, pressure, and heat removal \cite{Xia2017HCSG}. Marine nuclear systems may also be affected by platform-motion effects such as inclination, rolling, and heaving \cite{Hou2024HCOTSGInclination,Gong2023MarineReactorCFD,Xia2023SGSloshing}. The present work does not model these ocean-motion effects; instead, it focuses on propulsion-load-following dynamics under fixed-orientation operating conditions.

In this paper, an integrated dynamic model is used to study the load-following behavior of an SMR-powered all-electric ship. The model couples an SMR, once-through steam generator, turbine, synchronous generator, shipboard electrical network, asynchronous propulsion motor, and battery energy storage system (BESS). The SMR and turbine formulations are adopted from our prior dynamic modeling work \cite{Rad2026SMRLoadFollowing}, while the electrical subsystem is modeled in Simulink/Simscape Electrical. A propulsion motor torque reduction is applied as the load-following disturbance, and the resulting responses are observed throughout the nuclear-to-propulsion energy chain. The main contribution is a coupled system-level analysis showing how BESS support affects not only frequency regulation, but also the upstream steam cycle, reactor power, reactivity, and steam generator moving-boundary responses.

\section*{MODEL DESCRIPTION}

The integrated model combines a MATLAB implementation of the SMR and turbine dynamics with a Simulink/Simscape Electrical model of the shipboard electrical system. The SMR model includes reactor point kinetics, fuel and coolant temperature feedback, primary-loop thermal dynamics, a moving-boundary once-through steam generator, and a turbine valve model. The electrical model includes a synchronous generator, AC shipboard bus, static ship service load, asynchronous propulsion motor, and grid-connected BESS. MATLAB, Simulink, and Simscape Electrical are used as the simulation environment for coupling the nuclear, thermal, mechanical, and electrical subsystems \cite{MathWorksTools}. The detailed SMR and turbine formulation follows our prior study in \cite{Rad2026SMRLoadFollowing}; only the main coupling relationships are summarized here. The main parameters used for the integrated SMR-powered all-electric ship model are summarized in Table~\ref{tab:main_parameters}.

\begin{table}[htb!]
\centering
\caption{Main parameters of the integrated SMR-powered all-electric ship model.}
\label{tab:main_parameters}
\small
\setlength{\tabcolsep}{4pt}
\renewcommand{\arraystretch}{1.05}
\begin{tabular}{lll}
\hline
\textbf{Subsystem} & \textbf{Parameter} & \textbf{Value} \\
\hline
SMR & Rated thermal power, $P_{\mathrm{th},0}$ & $160~\mathrm{MW}$ \\
Steam cycle & Steam pressure, $p_s$ & $3.448~\mathrm{MPa}$ \\
Steam cycle & Rated steam flow, $\dot{m}_{s,0}$ & $67.07~\mathrm{kg\,s^{-1}}$ \\
Steam generator & Total tube length, $L_{\mathrm{SG}}$ & $24.2~\mathrm{m}$ \\
Generator & Rating and frequency & $50~\mathrm{MVA}$, $60~\mathrm{Hz}$ \\
Generator & Inertia constant, $H_G$ & $6~\mathrm{s}$ \\
Propulsion motor & Machine type and rating & Asynchronous, $40~\mathrm{MVA}$ \\
Static load & Active and reactive power & $10~\mathrm{MW}$, $5~\mathrm{MVAr}$ \\
BESS & Power and energy rating & $25~\mathrm{MW}$, $2~\mathrm{MWh}$ \\
BESS & Efficiency and SOC limits & $96\%$, $10$--$90\%$ \\
\hline
\end{tabular}
\end{table}

\subsection*{SMR and Turbine Model}

The normalized neutron flux, $\phi$, is used as the reactor power state, with $\phi=1$ corresponding to rated thermal power. The reactor thermal power is calculated as:
\begin{equation}
P_{\mathrm{th}}=\phi \times P_{\mathrm{th},0}.
\end{equation}
Reactivity feedback is represented using external reactivity and negative fuel and coolant temperature feedback:
\begin{equation}
\rho =
\rho_{\mathrm{ext}}
+\alpha_f \times (T_f-T_{f,0})
+\alpha_c \times (T_c-T_{c,0}),
\end{equation}
where $T_f$ is the fuel temperature, $T_c$ is the representative primary coolant temperature, and $\alpha_f$ and $\alpha_c$ are the fuel and coolant temperature reactivity coefficients, respectively. This feedback allows reactor power to respond to changes in heat removal from the steam generator during propulsion-load transients.

The steam generator is represented as a once-through moving-boundary heat exchanger with three secondary-side regions: subcooled liquid, saturated boiling flow, and superheated steam. The corresponding region lengths are $L_1$, $L_2$, and $L_3$, constrained by:
\begin{equation}
L_1+L_2+L_3=L_{\mathrm{SG}}.
\end{equation}
This representation captures the redistribution of the internal steam generator heat transfer regions as steam flow, secondary pressure, and heat removal change during load following.

The steam flow to the turbine is regulated through a simplified turbine-valve model. The valve command is passed through a servomotor with rate limits to represent finite turbine actuation speed. The turbine steam flow is approximated as:
\begin{equation}
\dot{m}_s=A_v \times p_s,
\end{equation}
where $A_v$ is the effective valve-flow coefficient and $p_s$ is the secondary-side steam pressure. The turbine mechanical power supplied to the synchronous generator is:
\begin{equation}
P_T=\eta_T \times \dot{m}_s \times \Delta h_T,
\end{equation}
where $\eta_T$ is the turbine efficiency and $\Delta h_T$ is the turbine enthalpy drop. Thus, electrical load changes do not instantaneously translate into turbine power changes; the response is mediated by the governor, valve servomotor, steam-generator pressure dynamics, and SMR thermal feedback.

\subsection*{Electrical Network, Propulsion Motor, and BESS}

The turbine mechanical power is supplied to the synchronous generator as $P_{m,G}=P_T$. The generator feeds an isolated AC bus connected to the propulsion motor, aggregated static load, and BESS. The key generator electromechanical response is governed by the mismatch between mechanical input power and electrical output power:
\begin{equation}
2H_G S_{G,\mathrm{base}} \times \frac{d\Delta \omega}{dt}
=
P_{m,G}
-
P_{e,G}
-
K_{D,G}S_{G,\mathrm{base}}\times\Delta\omega,
\end{equation}
where $\Delta\omega$ is the rotor-speed deviation, $H_G$ is the inertia constant, $S_{G,\mathrm{base}}$ is the generator base power, $P_{m,G}$ is the generator mechanical input power, $P_{e,G}$ is the generator electrical power, and $K_{D,G}$ is the damping coefficient. The system frequency is:
\begin{equation}
f=f_0 \times (1+\Delta\omega),
\end{equation}
with $f_0=60~\mathrm{Hz}$. The propulsion system is represented by a squirrel-cage asynchronous motor. A change in motor torque demand modifies motor speed, slip, electromagnetic torque, and electrical power drawn from the bus. At the system level, the active-power balance is:
\begin{equation}
P_{e,G}+P_{\mathrm{BESS}}
=
P_{\mathrm{motor}}+P_{\mathrm{static}}+P_{\mathrm{loss}},
\end{equation}
where positive $P_{\mathrm{BESS}}$ denotes discharge into the AC bus and negative $P_{\mathrm{BESS}}$ denotes charging or power absorption.

\subsection*{Frequency Control Strategy}

Frequency control is provided by two coordinated layers: the turbine governor controller and the BESS active power controller. The turbine governor uses generator speed error to adjust the turbine valve command through a proportional-integral controller and valve servomotor. This provides the slower sustained load-following response of the SMR-turbine generator.

The BESS provides faster frequency support using the measured frequency deviation,
\begin{equation}
\Delta f=f-f_0.
\end{equation}
The conceptual active power command is:
\begin{equation}
P_{\mathrm{BESS}}^{\mathrm{ref}}
=
-\left(
K_{f,B}\Delta f
+
K_{D,B}\frac{d\Delta f}{dt}
\right),
\end{equation}
where $K_{f,B}$ and $K_{D,B}$ are the frequency-deviation and frequency-rate feedback gains, respectively. Therefore, when frequency rises after a propulsion-load reduction, the BESS absorbs active power; when frequency falls after a load increase, the BESS injects active power. In implementation, the BESS command is passed through filtering, deadband logic, saturation, rate limits, and state of charge (SOC) constraints. This separates the fast frequency-arrest function of the BESS from the slower nuclear plant load-following function of the SMR turbine generator.


\section*{RESULTS AND DISCUSSION}

\subsection{Propulsion Load-Reduction Scenario}

The load-following response of the integrated SMR-powered all-electric ship system is evaluated using a propulsion-load reduction. The mechanical torque command of the propulsion motor is initially set to $0.9785~\mathrm{pu}$ and is reduced to $0.8~\mathrm{pu}$ between $t=100~\mathrm{s}$ and $t=105~\mathrm{s}$. This represents an approximately $18.2\%$ reduction in propulsion torque demand over $5~\mathrm{s}$. The same disturbance is applied to two configurations: the SMR-powered system without BESS support and the SMR-powered system with BESS support.

\begin{figure}[htb!]
\centering
\begin{subfigure}{0.48\linewidth}
    \centering
    \includegraphics[width=\linewidth]{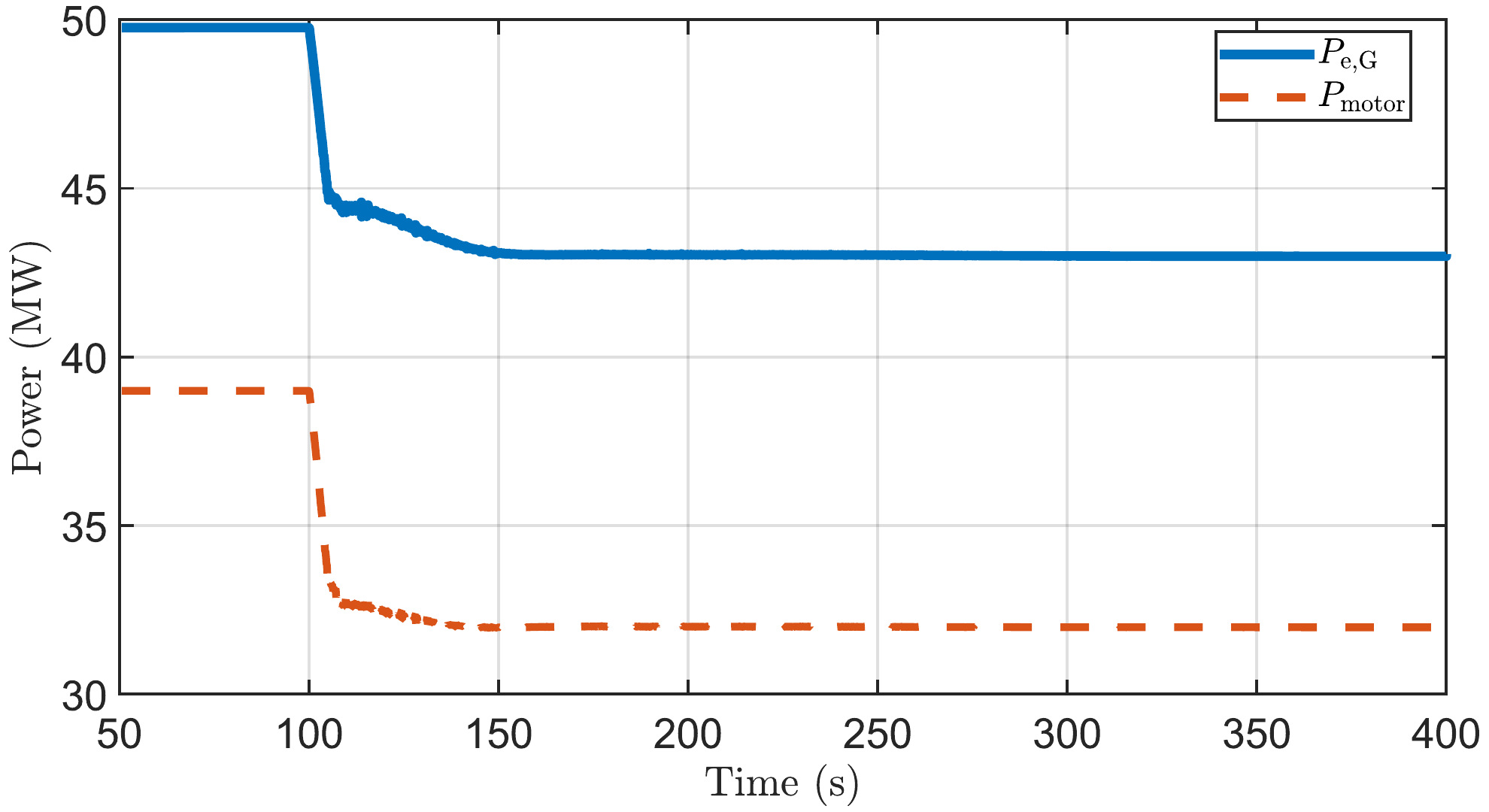}
    \caption{SMR alone}
    \label{fig:power_smr_alone}
\end{subfigure}
\hfill
\begin{subfigure}{0.48\linewidth}
    \centering
    \includegraphics[width=\linewidth]{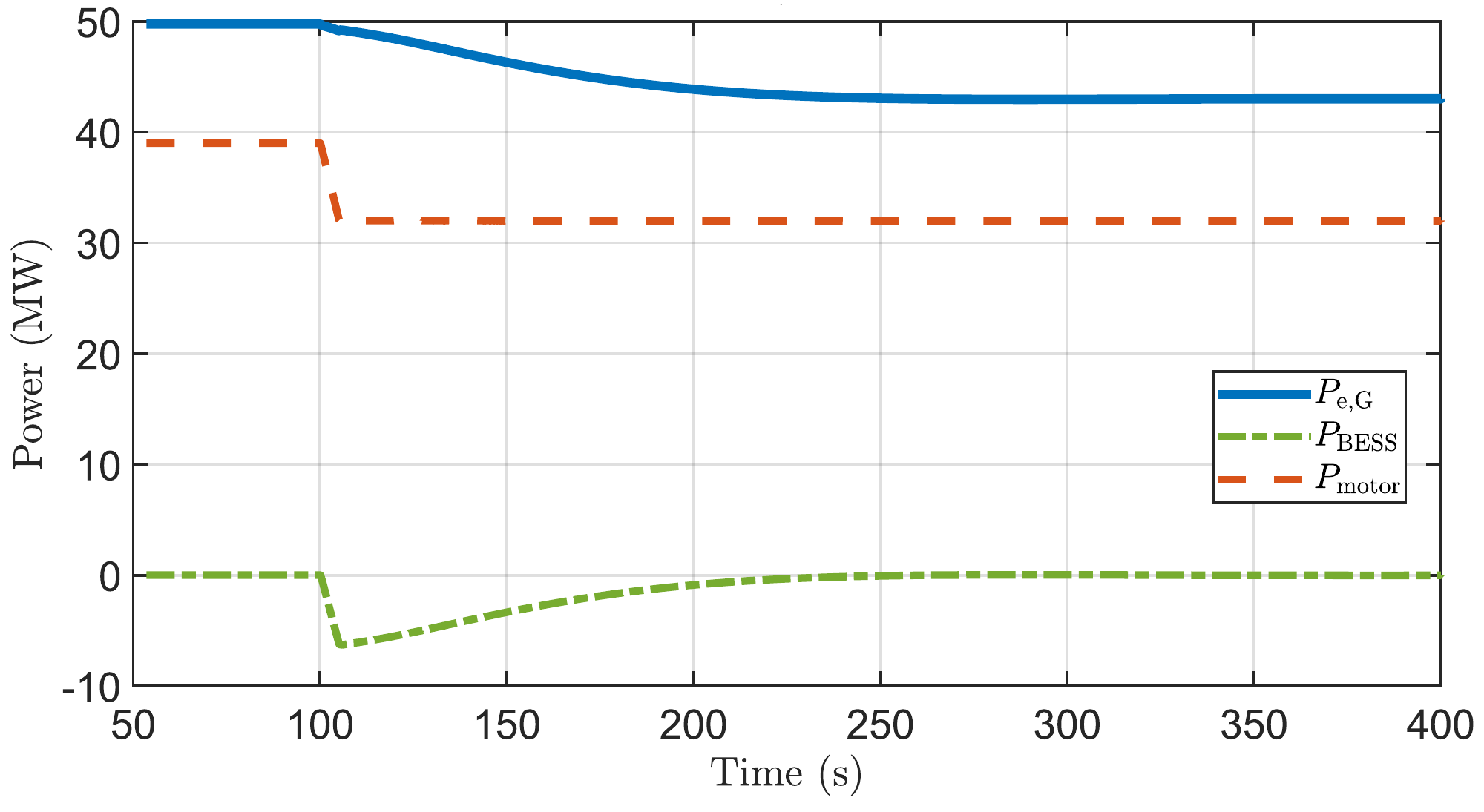}
    \caption{SMR + BESS}
    \label{fig:power_smr_bess}
\end{subfigure}
\caption{Electrical power response to propulsion-load reduction.}
\label{fig:power_comparison}
\end{figure}

Figure~\ref{fig:power_comparison} shows the electrical power response for both configurations. In the SMR-alone case, the propulsion motor power decreases rapidly after the torque reduction, and the generator electrical power also decreases as the system moves toward the new load condition. The generator power remains higher than the motor power because the generator also supplies the static ship-service load and electrical losses. The transient difference between generation and demand is reflected in the frequency response and is handled only by the generator inertia and the turbine-governor dynamics.

In the SMR+BESS case, the BESS absorbs power immediately after the propulsion-load reduction. Because the load decreases faster than the SMR-turbine generator can reduce its mechanical power, the battery active power becomes negative, indicating charging. This allows the BESS to absorb part of the temporary excess generation and smooth the power imbalance seen by the electrical network. As the turbine-governor and SMR system approach the new operating condition, the battery power gradually returns toward zero. Therefore, the BESS provides fast transient power balancing, while the SMR-turbine generator provides the slower sustained load-following response.

\begin{figure}[htb!]
\centering
\begin{subfigure}{0.48\linewidth}
    \centering
    \includegraphics[width=\linewidth]{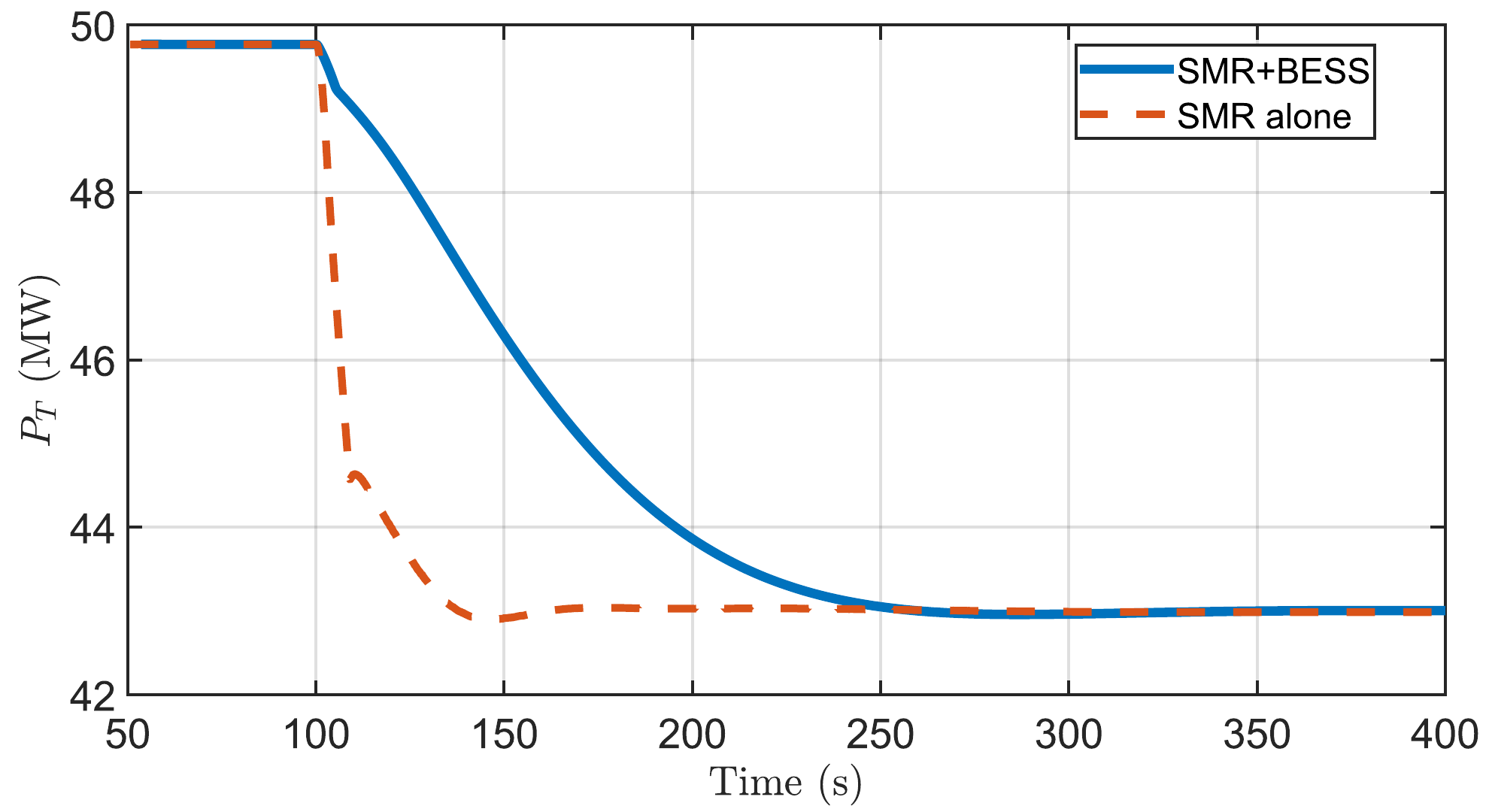}
    \caption{Turbine mechanical power}
    \label{fig:turbine_power_comparison}
\end{subfigure}
\hfill
\begin{subfigure}{0.48\linewidth}
    \centering
    \includegraphics[width=\linewidth]{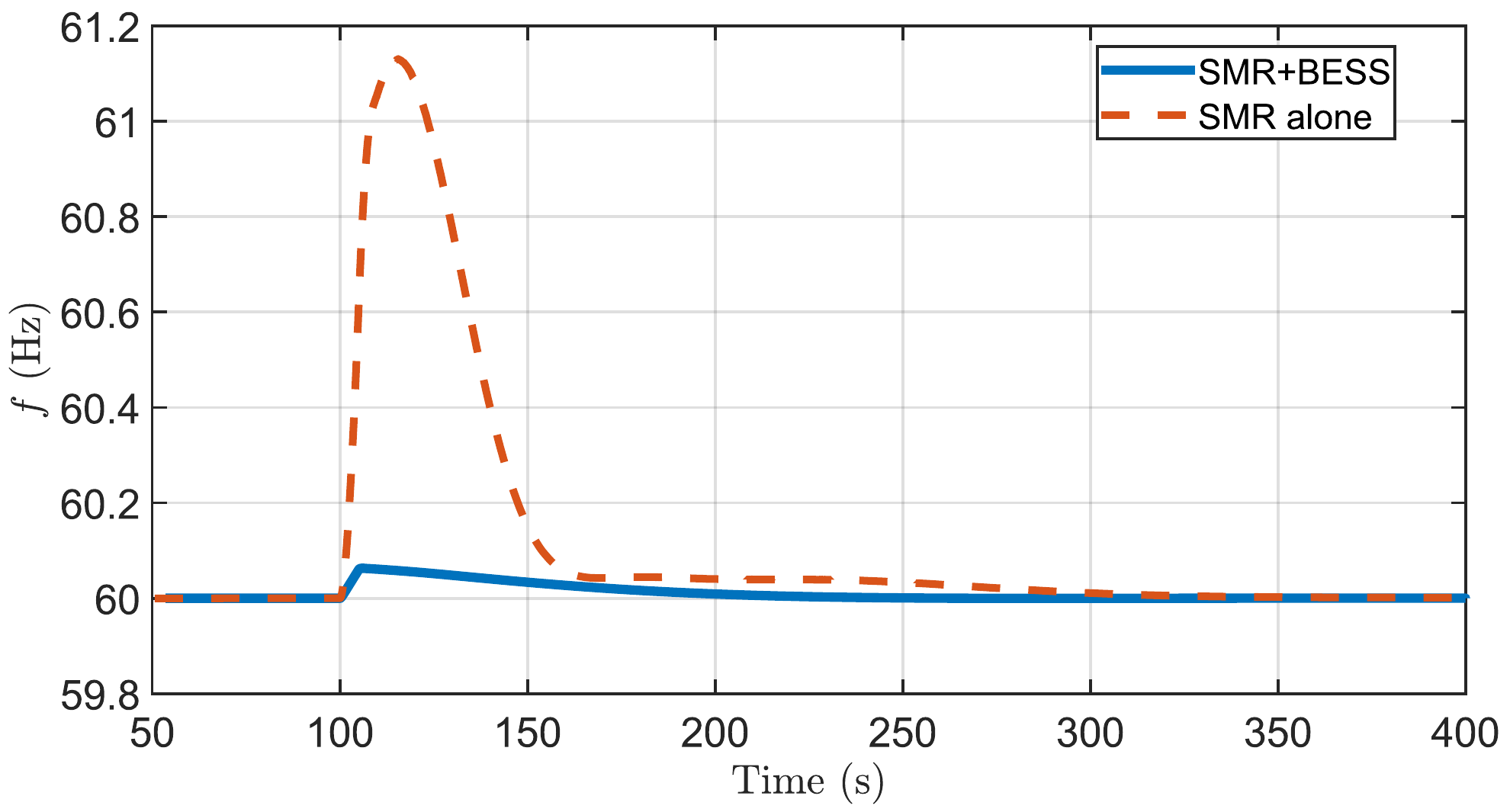}
    \caption{Frequency}
    \label{fig:frequency_comparison}
\end{subfigure}
\caption{Comparison of turbine mechanical power and frequency following propulsion-load reduction for the SMR-alone and SMR+BESS configurations.}
\label{fig:turbine_frequency_comparison}
\end{figure}

Figure~\ref{fig:turbine_frequency_comparison} compares the turbine mechanical power and generator frequency for the two configurations. In the SMR-alone case, the propulsion-load reduction produces a rapid surplus of mechanical generation relative to electrical demand. As a result, the system frequency increases sharply, reaching a peak value of approximately $61.1~\mathrm{Hz}$. The turbine-governor then reduces the turbine mechanical power rapidly as it attempts to restore the power balance.

When the BESS is included, the frequency deviation is significantly reduced. Immediately after the load reduction, the BESS absorbs part of the excess electrical power, which limits the over-frequency transient. As a result, the peak frequency remains close to nominal, increasing only to approximately $60.06~\mathrm{Hz}$. The turbine mechanical power also changes more gradually in the SMR+BESS case, indicating that the battery absorbs the fast electrical mismatch while the SMR-turbine system follows the new load condition on a slower time scale.

\subsection{Steam-Cycle Response of the SMR System}

The steam-cycle response is evaluated using three variables: turbine steam mass flow rate, secondary-side steam-generator pressure, and feedwater mass flow rate. These variables show how the propulsion-load reduction propagates upstream through the turbine valve, steam generator, and feedwater control system.

\begin{figure}[htb!]
\centering
\begin{subfigure}{0.48\linewidth}
    \centering
    \includegraphics[width=\linewidth]{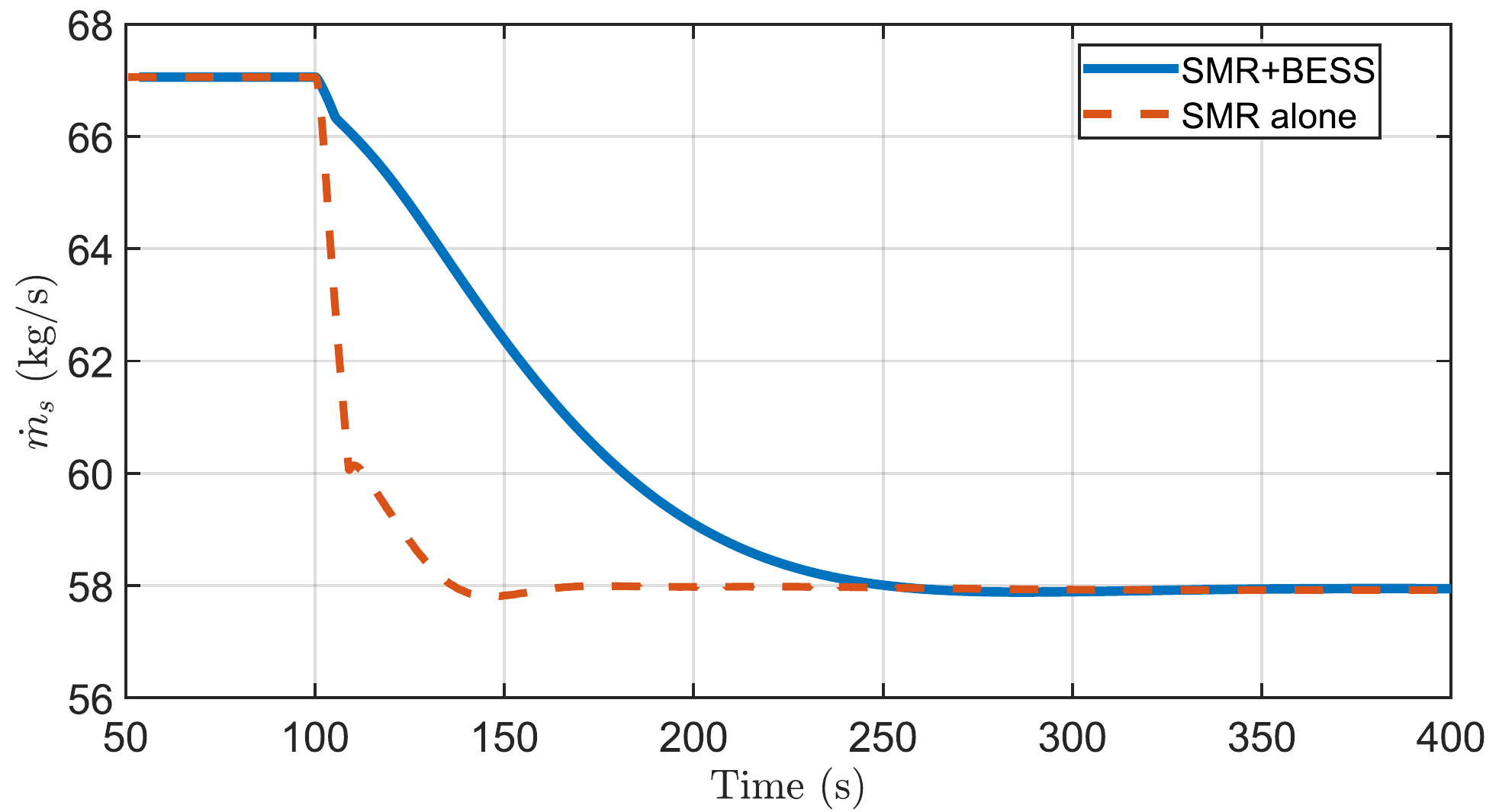}
    \caption{Steam mass flow rate}
    \label{fig:steam_mass_flow_comparison}
\end{subfigure}
\hfill
\begin{subfigure}{0.48\linewidth}
    \centering
    \includegraphics[width=\linewidth]{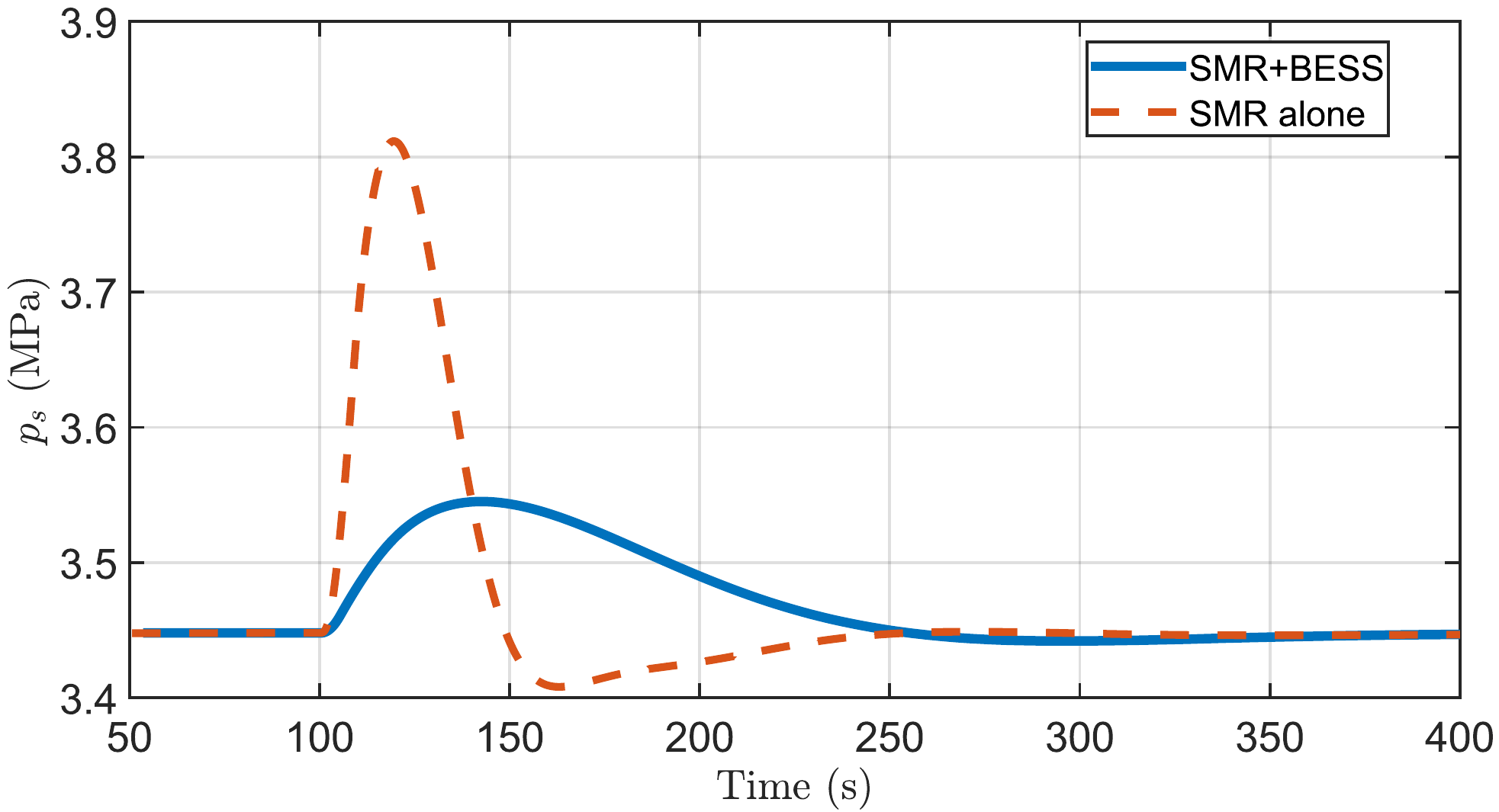}
    \caption{Steam generator secondary pressure}
    \label{fig:steam_pressure_comparison}
\end{subfigure}
\hfill
\begin{subfigure}{0.48\linewidth}
    \centering
    \includegraphics[width=\linewidth]{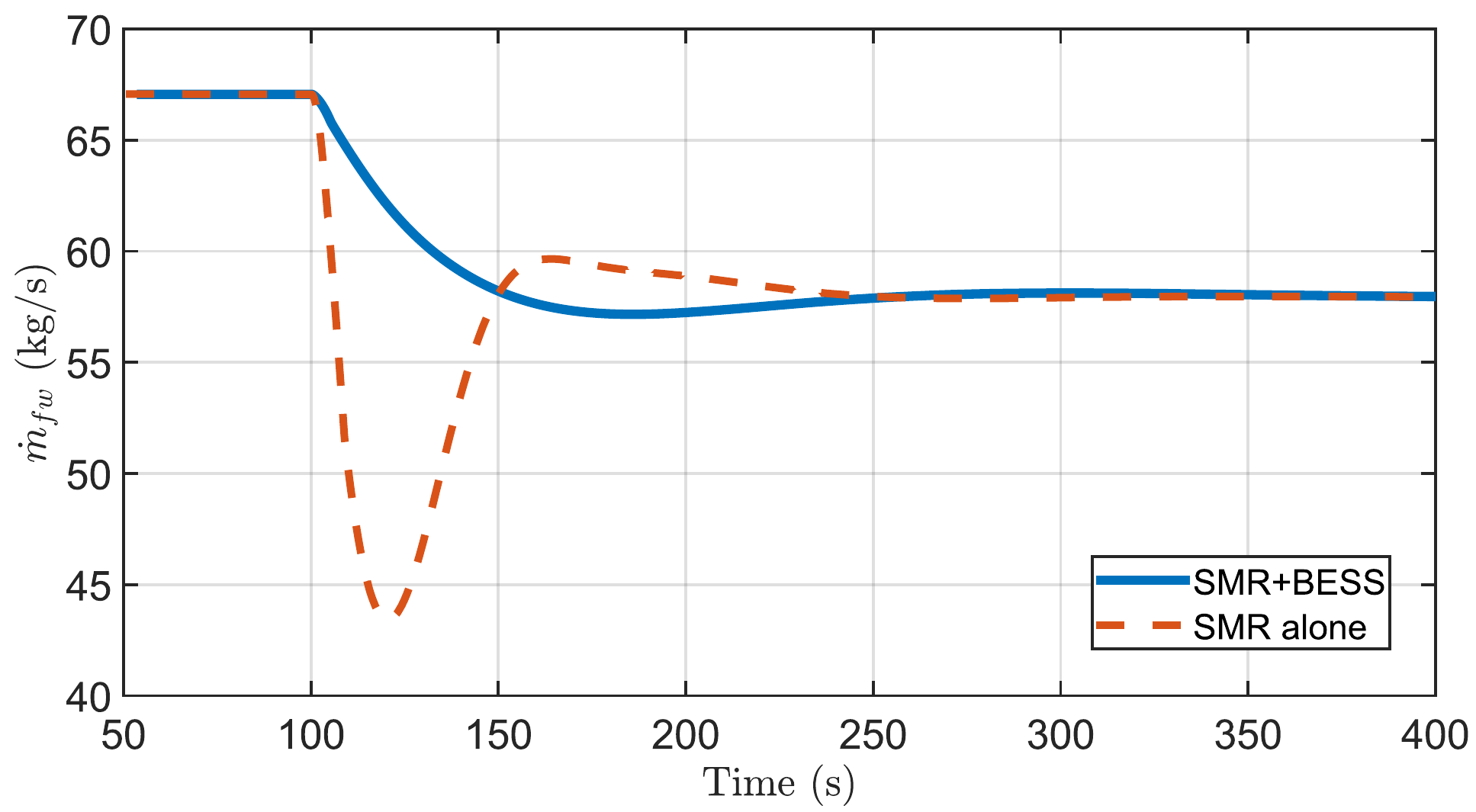}
    \caption{Feedwater mass flow rate}
    \label{fig:feedwater_mass_flow_comparison}
\end{subfigure}

\caption{Steam-cycle response following propulsion-load reduction for the SMR-alone and SMR+BESS configurations.}
\label{fig:steam_cycle_response}
\end{figure}

As shown in Fig.~\ref{fig:steam_cycle_response}, the SMR-alone case exhibits a sharper steam-cycle transient after the propulsion-load reduction. The turbine valve closes rapidly to reduce turbine mechanical power, which produces a sharp decrease in steam mass flow rate. Because steam outflow is restricted, mass inventory temporarily accumulates in the steam generator, causing a rise in the secondary-side pressure. The feedwater controller then reduces the feedwater mass flow rate to restore the secondary-side pressure and mass balance.

When the BESS is included, the same propulsion-load reduction produces a smoother steam-cycle response. The BESS absorbs part of the fast excess electrical power, reducing the immediate change required from the turbine-generator system. As a result, the steam mass flow rate decreases more gradually, the steam-generator pressure excursion is smaller, and the feedwater-flow adjustment is less severe. This result shows that fast electrical energy storage can indirectly reduce upstream thermal-hydraulic transients in the SMR steam cycle.

\subsection{Primary-Loop and Reactor Power Response}

The upstream reactor response is evaluated using the average primary coolant temperature, total reactivity, and reactor thermal power. These variables show how the steam-cycle disturbance affects the primary-side thermal state and the neutronic response of the SMR. Since the propulsion-load reduction decreases the required turbine power, the steam flow and heat-removal conditions in the steam generator change, which then affects the primary coolant temperature, reactivity feedback, and reactor thermal power.

\begin{figure}[htb!]
\centering

\begin{subfigure}{0.48\linewidth}
    \centering
    \includegraphics[width=\linewidth]{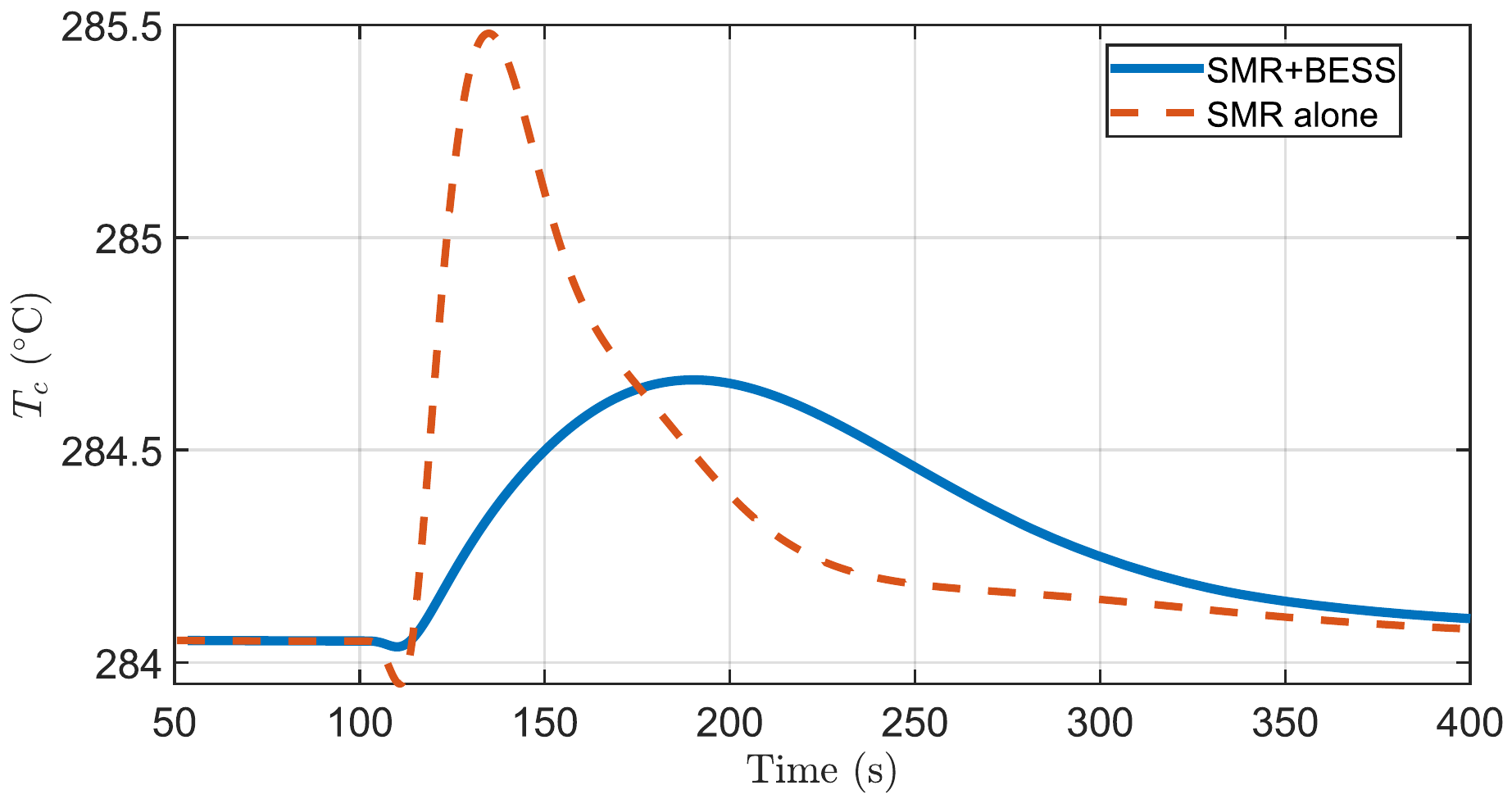}
    \caption{Primary coolant temperature}
    \label{fig:primary_coolant_temperature_comparison}
\end{subfigure}
\hfill
\begin{subfigure}{0.48\linewidth}
    \centering
    \includegraphics[width=\linewidth]{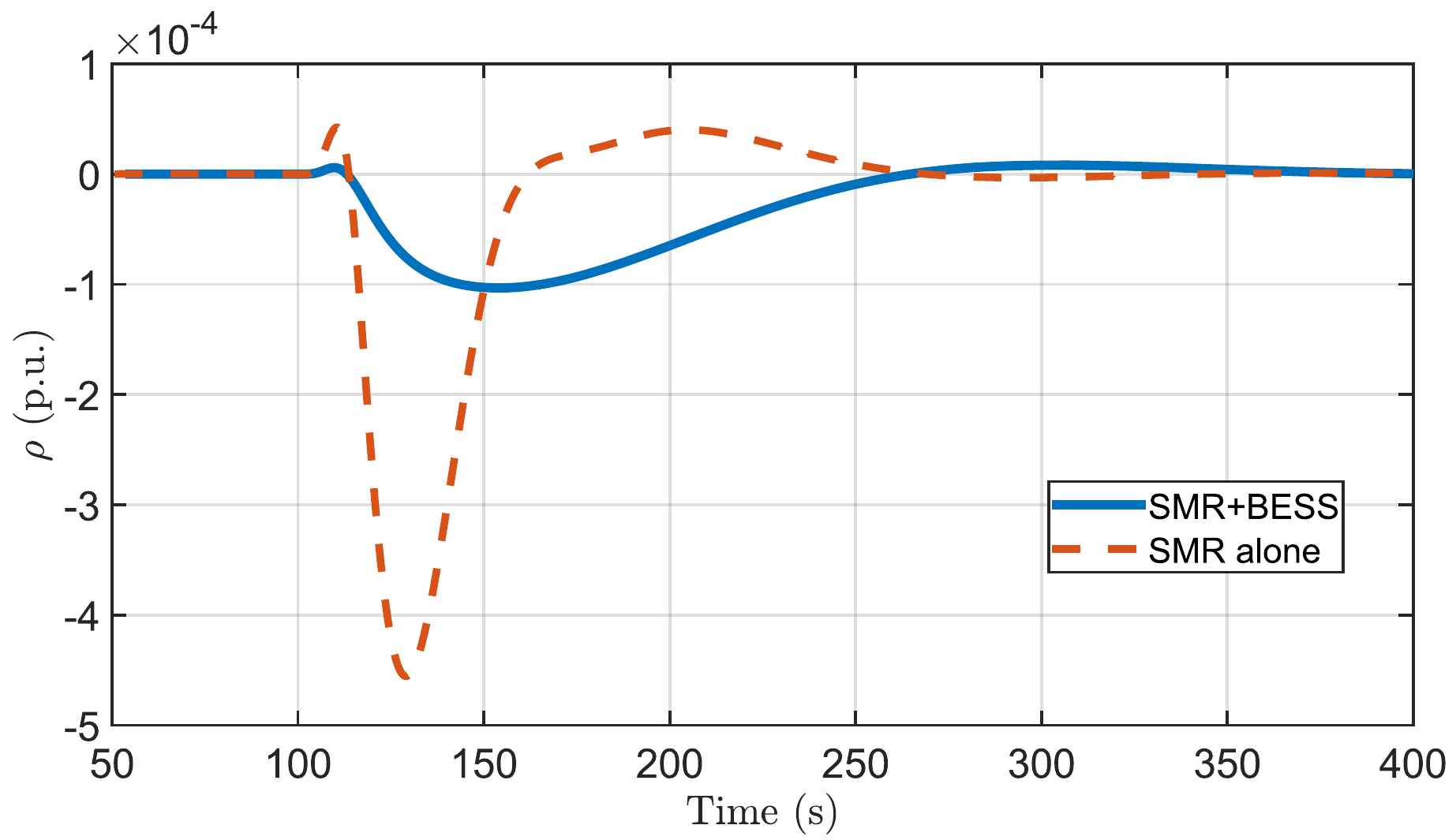}
    \caption{Total reactivity}
    \label{fig:reactivity_comparison}
\end{subfigure}
\hfill
\begin{subfigure}{0.48\linewidth}
    \centering
    \includegraphics[width=\linewidth]{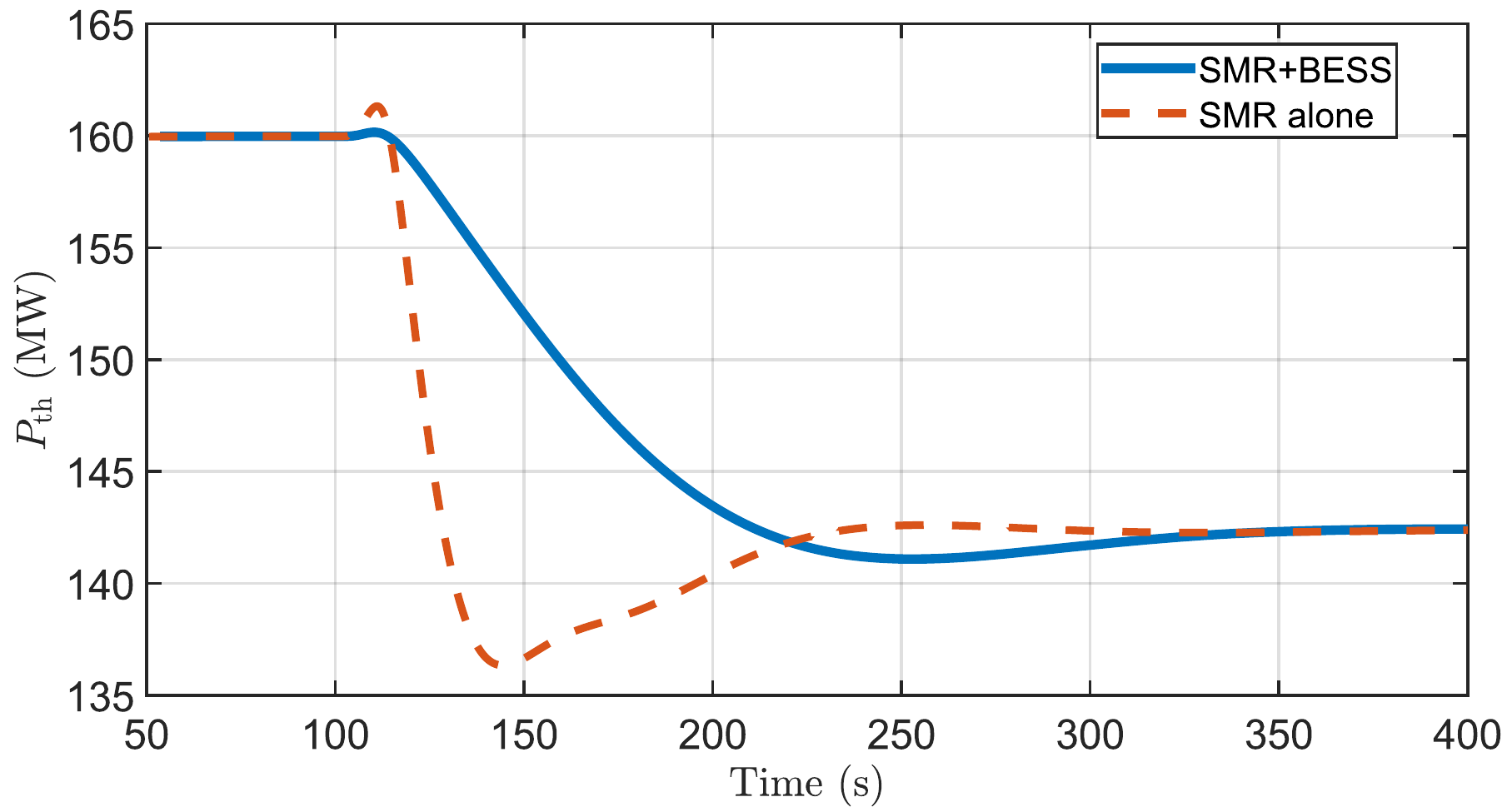}
    \caption{Reactor thermal power}
    \label{fig:reactor_thermal_power_comparison}
\end{subfigure}

\caption{Primary-loop and reactor-power response after propulsion-load reduction.}
\label{fig:primary_reactor_response}
\end{figure}

As shown in Fig.~\ref{fig:primary_reactor_response}, the SMR-alone case produces a sharper primary-side thermal and neutronic transient. After the propulsion-load reduction, the turbine valve closes rapidly and the steam mass flow rate decreases. This changes the heat-removal condition in the steam generator and causes the primary coolant temperature to rise. The increase in coolant temperature contributes negative reactivity through the temperature-feedback mechanism, which reduces the reactor thermal power.

When the BESS is included, the reactor-side response becomes smoother. The BESS absorbs part of the fast excess electrical power after the load reduction, reducing the immediate change required from the turbine-generator system. As a result, the steam-cycle disturbance is softened, the primary coolant temperature rises more gradually, the negative reactivity excursion is reduced, and the reactor thermal power decreases more smoothly toward the new operating condition. This confirms that the BESS does not directly control the reactor, but it indirectly moderates the reactor thermal response by reducing the severity of the upstream steam-cycle transient.

\subsection{Steam Generator Moving-Boundary Response}

The final set of SMR-side variables considered in this study is the steam generator moving-boundary response. The once-through steam generator is represented by three secondary-side regions: the subcooled region, $L_1$, the saturated region, $L_2$, and the superheated region, $L_3$. These region lengths show how the internal heat-transfer structure of the steam generator changes during the propulsion-load reduction.

\begin{figure}[htb!]
\centering

\begin{subfigure}{0.48\linewidth}
    \centering
    \includegraphics[width=\linewidth]{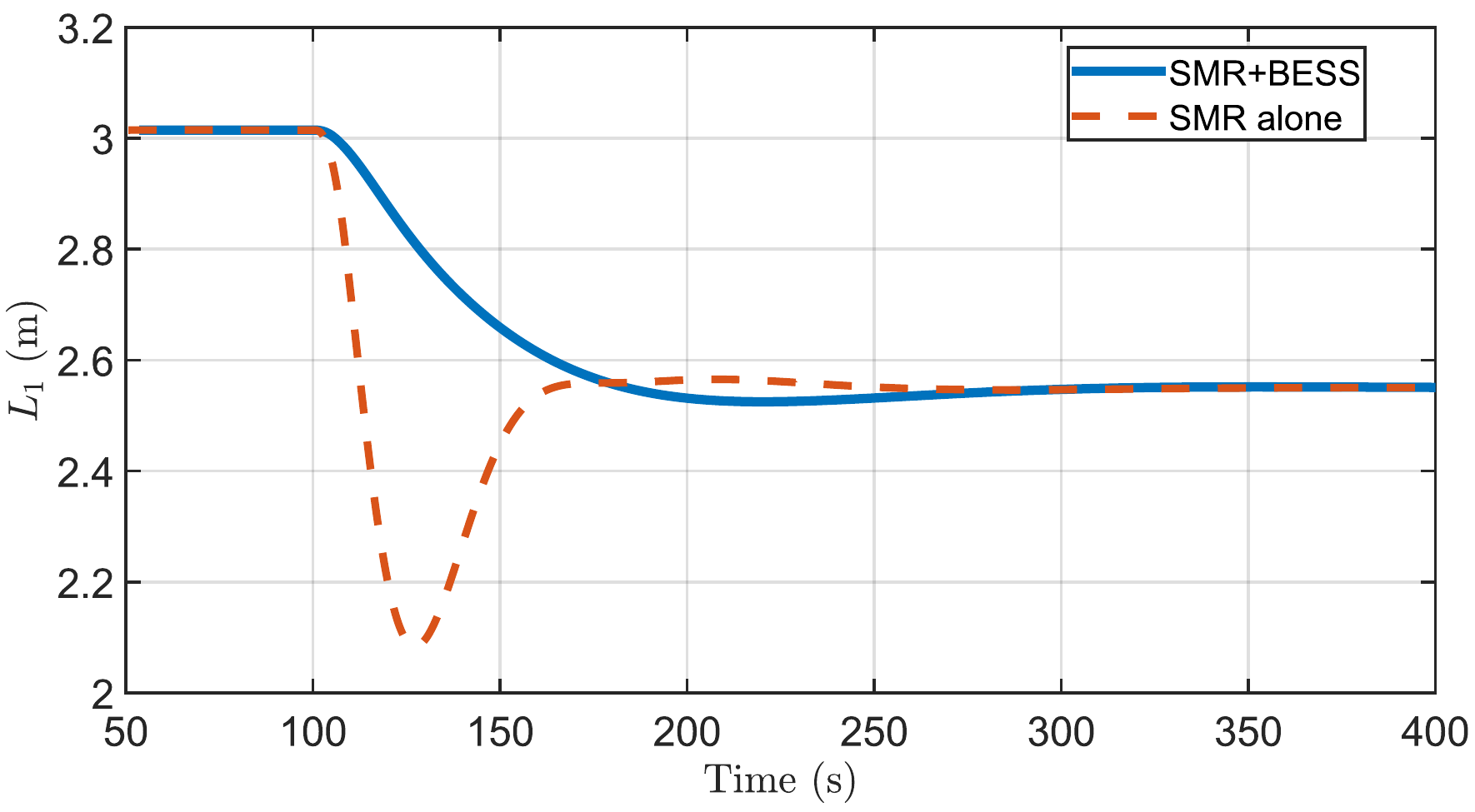}
    \caption{Subcooled region, $L_1$}
    \label{fig:L1_comparison}
\end{subfigure}
\hfill
\begin{subfigure}{0.48\linewidth}
    \centering
    \includegraphics[width=\linewidth]{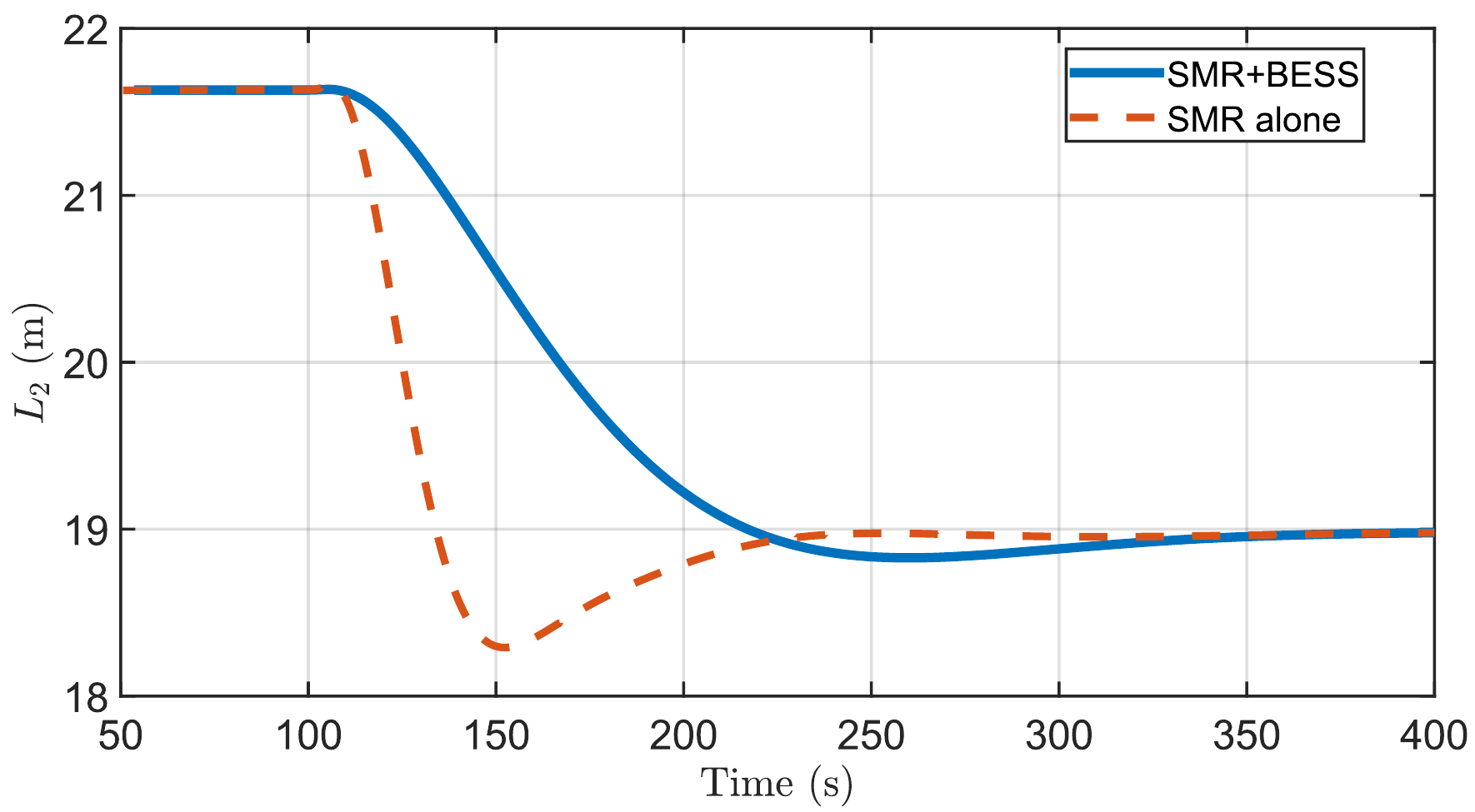}
    \caption{Saturated region, $L_2$}
    \label{fig:L2_comparison}
\end{subfigure}
\hfill
\begin{subfigure}{0.48\linewidth}
    \centering
    \includegraphics[width=\linewidth]{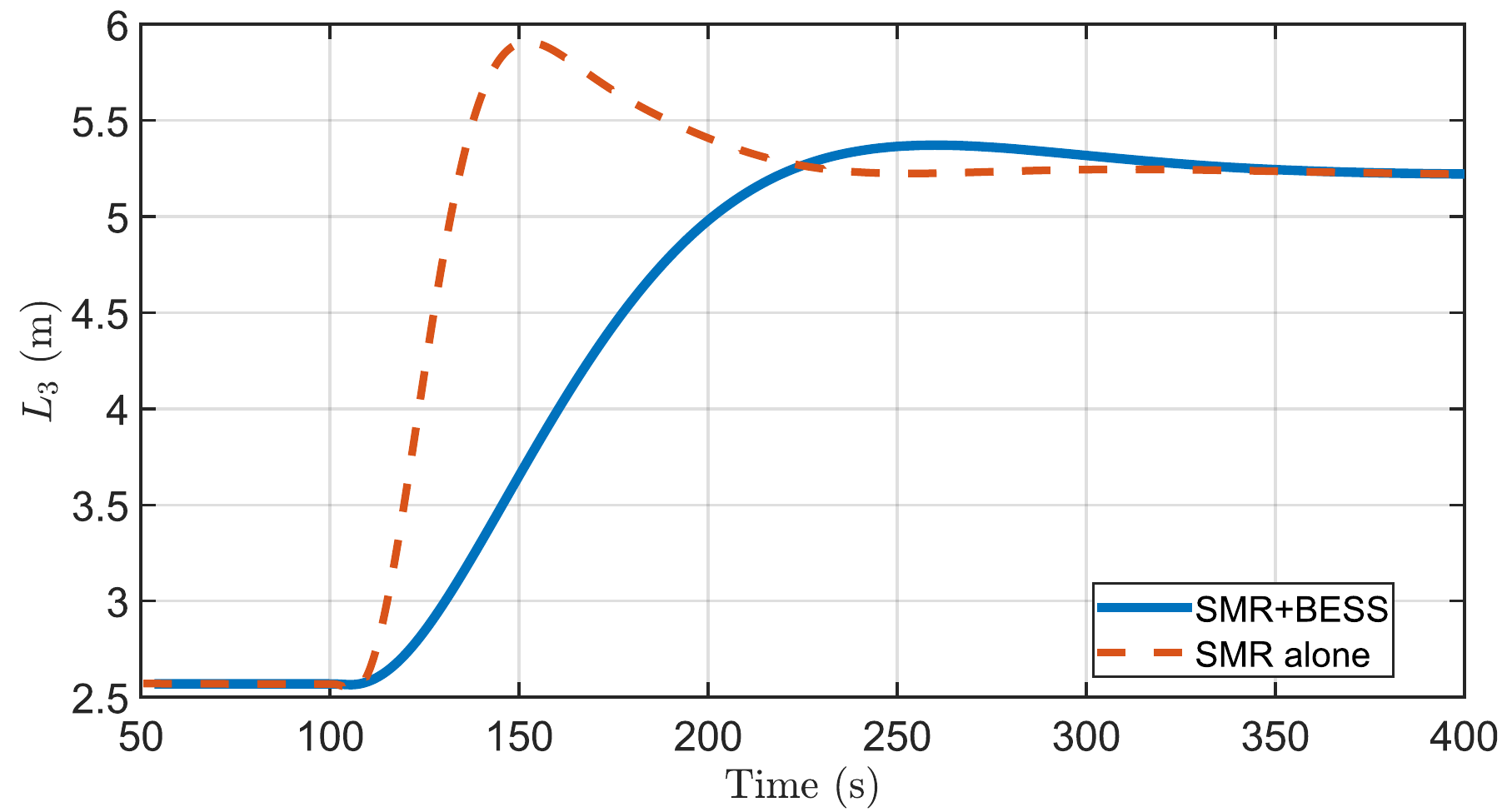}
    \caption{Superheated region, $L_3$}
    \label{fig:L3_comparison}
\end{subfigure}

\caption{Steam generator region-length response after propulsion-load reduction.}

\label{fig:sg_boundary_response}
\end{figure}

As shown in Fig.~\ref{fig:sg_boundary_response}, the propulsion-load reduction changes the internal phase-region distribution of the once-through steam generator. In the SMR-alone case, the rapid decrease in steam flow and the sharp pressure transient produce a strong redistribution of the steam-generator regions. The subcooled and saturated regions, $L_1$ and $L_2$, decrease sharply, while the superheated region, $L_3$, increases rapidly before the system gradually approaches the new operating condition.

When the BESS is included, the moving-boundary response becomes smoother. Because the BESS absorbs part of the fast electrical power surplus, the turbine and steam-cycle variables change more gradually. As a result, the changes in $L_1$, $L_2$, and $L_3$ are less abrupt than in the SMR-alone case. This result further confirms that fast electrical storage can indirectly reduce thermal-hydraulic transients in the steam generator by smoothing the upstream effect of the propulsion-load reduction.

\section*{CONCLUSIONS}

This paper presented an integrated dynamic load-following analysis of an SMR-powered all-electric ship. The model couples an integral pressurized water reactor, a once-through moving-boundary steam generator, a turbine generator system, an AC shipboard electrical network, an asynchronous propulsion motor, and a battery energy storage system. A propulsion-load reduction was applied through the motor torque command, and two configurations were compared: the SMR-powered system alone and the SMR-powered system with BESS support.

The results show that the propulsion-load reduction first appears as a fast electrical disturbance at the shipboard bus. Without BESS support, the sudden decrease in motor power creates a temporary surplus between mechanical generation and electrical demand, producing a large over-frequency transient and a rapid reduction in turbine mechanical power. When the BESS is included, the battery absorbs excess electrical power during the load reduction, which significantly reduces the frequency overshoot and smooths the generator and turbine response.

The comparison also shows that BESS support affects more than the electrical subsystem. By reducing the immediate turbine-power adjustment required after the load reduction, the BESS indirectly softens the upstream steam-cycle and reactor response. The SMR+BESS case produces smoother changes in steam mass flow rate, steam-generator pressure, feedwater flow, primary coolant temperature, reactivity, reactor thermal power, and steam-generator region lengths. These results demonstrate that fast electrical energy storage can reduce the severity of thermal-hydraulic and neutronic transients in an SMR-powered all-electric ship, even though the BESS does not directly control the reactor.

Overall, the study highlights the importance of evaluating nuclear-powered electric propulsion systems as coupled nuclear, thermal, mechanical, and electrical systems. The results suggest that BESS integration can improve short-term frequency regulation while allowing the SMR and steam cycle to follow propulsion-load changes on a slower time scale. Future work will extend the model to include additional propulsion-load profiles, coordinated energy management strategies, detailed BESS sizing, and marine-motion effects such as inclination, rolling, and heaving.

\section*{ACKNOWLEDGMENTS}
This work was supported by the U.S. Department of Energy under Awards DE-NE0009296 and DE-NE0009643.

\bibliographystyle{IEEEtran}
\bibliography{references}









\end{document}